\documentstyle[epsf,aps,pre,psfig,floats]{revtex}

\begin{document}

\title{Measurements of Particle Dynamics in Slow, Dense Granular Couette Flow}
\author{Daniel M. Mueth}
\address{The James Franck Institute and Department of Physics\\
        The University of Chicago\\
        5640 S. Ellis Ave., Chicago, IL 60637}
\maketitle

\bibliographystyle{prsty}

\begin{abstract}
Experimental measurements of particle dynamics on the lower surface of
a 3D Couette cell containing monodisperse spheres are reported.  The
average radial density and velocity profiles are similar to those
previously measured within the bulk and on the lower surface of the 3D
cell filled with mustard seeds.  Observations of the evolution of
particle velocities over time reveal distinct motion events, intervals
where previously stationary particles move for a short duration before
jamming again. The cross-correlation between the velocities of two
particles at a given distance $r$ from the moving wall reveals a
characteristic lengthscale over which the particles are correlated.
The autocorrelation of a single particle's velocity reveals a
characteristic timescale $\tau$ which decreases with distance from the
inner moving wall. This may be attributed to the increasing rarity at
which the discrete motion events occur and the reduced duration of
those events at large $r$.  The relationship between the RMS azimuthal
velocity fluctuations, $\delta v_\theta(r)$, and average shear rate,
$\dot\gamma(r)$, was found to be $\delta v_\theta \propto
\dot\gamma^\alpha$ with $\alpha = 0.52 \pm 0.04$.  These observations
are compared with other recent experiments and with the modified
hydrodynamic model recently introduced by Bocquet et al.


\end{abstract}

\section*{Introduction}

The detailed understanding of slow flow in dense granular systems has
remained one of the central challenges in the field of granular
materials~\cite{jaeger96}.  While fast dilute granular flows are
fluid-like and can be described well by granular kinetic
theory~\cite{bagnold54,ogawa80,jenkins83}, slow flows at high packing
fraction preserve many of the complex properties of static granular
packs and may be more accurately described as slow, plastic
deformation of a metastable granular solid than flow of a fluid.
Recent work by Howell and Behringer~\cite{howell97b,howell99} showed
that many of the intriguing properties of granular solids, such as the
broad distribution of stresses and the presence of ``force chains''
which focus stresses along paths of many connected particles, play a
crucial role.  Direct visualization revealed that the flow is
intermittant in time and that correlations, both in time and in space,
exist~\cite{videoofcorrelations}.  These are seen as intervals over
which one or more particles in a region become unjammed, move for a
short time, and then become jammed again.  Such properties of dense
granular flow are reminiscent of behavior seen in glasses, dense
colloidal suspensions, and foams~\cite{liu98}.  Recent studies have
been successful at relating stresses in stationary bead packs with
those in glassy fluids~\cite{ohern01}, suggesting the two systems may
also have similar flow properties for high packing fractions and low
shearing rates. It is hoped that an understanding of dense granular
flow will provide insight into the properties of static granular packs
as well as the broader class of jammed systems.

From previous work a number of unresolved fundamental questions about
slow, dense granular shear flow emerge. Although videos and plots of
particle trajectories suggest that particle velocities are correlated
both in space and in time, these correlations have not been directly
measured or characterized, and their effect on the overall flow
behavior is unknown.  Previous experiments performed with various
granular materials~\cite{mueth00,howellprivcommun} have shown that the
microscopic properties of a material such as particle shape,
polydispersity, and surface friction manifest themselves in the
macroscopic flow behavior; however the details of how the microscopic
particle properties and dynamics influence the overall flow is not
fully understood.  A number of theories describe the flow by relying
on coarse-grained quantities, such as an effective temperature or
viscosity.  However, as we show here, rather than vary smoothly and
monotonically, these quantities change abruptly and even oscillate on
sub-particle lengthscales which may make it difficult to capture the
full behavior in a coarse-grained model.  The lack of a complete
experimental description of the system, especially on the microscopic
scale, has so far prevented the rigorous testing of any of these
models.

To resolve these issues, a detailed description of the particle
dynamics on the microscopic, ie. single particle level, is needed.
While a number of careful, high-resolution,
measurements~\cite{howell99,mueth00,khosropour97,veje99,losert00,karion00,bocquet01}
of the average macroscopic properties of slow, dense Couette flow have
been performed, a detailed description of the microscopic dynamics has
been lacking.  The goal of the experiments reported in this article is
to characterize the particle dynamics on both microscopic and
mesoscopic scales, and to measure both the average properties of the
material as well as the fluctuations about the mean.

To this end a systematic study of the dynamics of particles on the
lower surface of a three-dimensional (3D) Couette cell was performed.
The dynamics were captured with digital video, and individual
particles were tracked through time.  Accurate measurements of the
particle velocity fluctuations and correlations required
high-precision measurements of the particle positions in every video
frame.  In contrast to measurements of the time-averaged velocity or
density profile, where the effects of measurement noise is reduced by
sufficient averaging~\cite{mueth00}, high-bandwidth fluctuation
measurements are limited by the precision of each position
measurement.  For this reason, a number of improvements in the
position measurement technique were implemented and extensive
calibration experiments made it possible to mitigate the effects of
instrumental noise on the fluctuation and correlation measurements.

With these improvements, I was able to measure correlations in
particle velocities and to obtain the correlation length $\xi$ and
correlation time $\tau$.  The improved resolution also allowed me to
measure the distribution and amplitude of the velocity fluctuations
over a much larger range than was previously possible.  The ability to
determine, locally and simultaneously in the same system the packing
fraction, average velocity, and fluctuation amplitude allowed for a
direct comparison with granular kinetic or hydrodynamic models, such
as the recent model by Bocquet et al.~\cite{bocquet01}.

\section*{Background}

\begin{figure}
\centerline{
\psfig{file=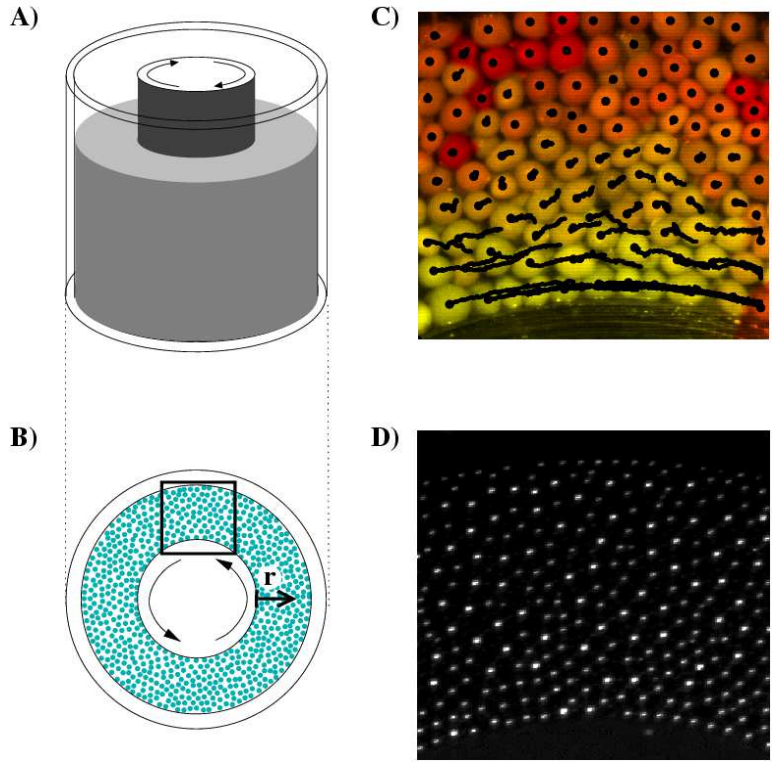,width=4.0in}}
\vspace{2ex}
\caption{ Sketch of the Couette cell and images of the lower surface.
(A) The Couette cell consists of two coaxial cylinders of radii 25.0
and 41.5~mm.  The gap between the cylinders is filled with a granular
material and the inner cylinder is rotated at constant angular
velocity.  (B) The motion of the particles near the lower surface is
recorded with a high speed digital camera at 250 frames per
second. The region imaged is represented by the rectangle. 
(C) The trajectory and average velocity of each particle over the
previous 200~ms (indicated by the particle shade: brighter for fast or
darker for slow) are shown for mustard seeds, obtained using particle
tracking techniques.  (D) An example image of the cell filled with
stainless steel bearing balls which are used for experiments
requiring high precision position measurements.}
\label{fig:sketch}
\end{figure}

Recently, several experiments on slow granular flows have been
performed in the Couette geometry, both in two-dimensional
(2D)~\cite{veje99,howell99} and
3D~\cite{khosropour97,mueth00,losert00,karion00,bocquet01} systems.
The Couette geometry (see Fig.~\ref{fig:sketch}A) with uniformly
rotating inner cylinder provides a steady-state flow which can be
maintained for very long times and has a symmetry for azimuthal
averaging of many quantities of interest.  The break-up of the system
into a shear band near the inner cylinder and an essentially stagnant
material near the outer cylinder furthermore provides a single system
in which velocities vary substantially through the cell allowing one
to view faster and slower shearing regions simultaneously.

Because dry granular materials are opaque, observations are either
confined to 2D Couette cells~\cite{veje99,howell99} and to the outer
surfaces of 3D cells~\cite{mueth00,losert00,karion00,bocquet01}, where
optical tracking of surface particles is possible, or non-invasive 3D
imaging tools must be employed~\cite{khosropour97,mueth00}. For 2D
systems and the surfaces of 3D systems, the motion is typically
recorded with video and each particle's motion is tracked in the
digitized frames using software.  This approach has the benefit of
providing detailed information about the particle dynamics. However,
it is unable to probe the bulk behavior of a 3D system. Furthermore,
packing fractions can only be determined where the particles are
guaranteed to lie in the imaging plane, such as along the lower
surface of the container.  Non-invasive 3D imaging techniques, such as
MRI and X-ray tomography, can be very effective at measuring certain
properties of a 3D system. X-ray tomography is very well suited for
measuring the time-averaged packing fraction~\cite{mueth00} and is a
valuable imaging
tool~\cite{mueth00,xrayimagingnote,seidler00}. However, determining
the precise particle positions in 3D is computationally expensive and
data acquisition rates are very slow, making 3D particle tracking
studies difficult. MRI has proven itself as a powerful tool for
measuring the time-averaged velocity inside 3D granular
flows~\cite{fukushima99,nakagawa93,ehrichs95} and has been used to
measure the average flow properties within the bulk in granular
Couette flow~\cite{mueth00}.  MRI has also been used to measure the
diffusion coefficient and correlation times for flowing granular
material~\cite{seymour00,caprihan00}.

\begin{figure}
\centerline{
\psfig{file=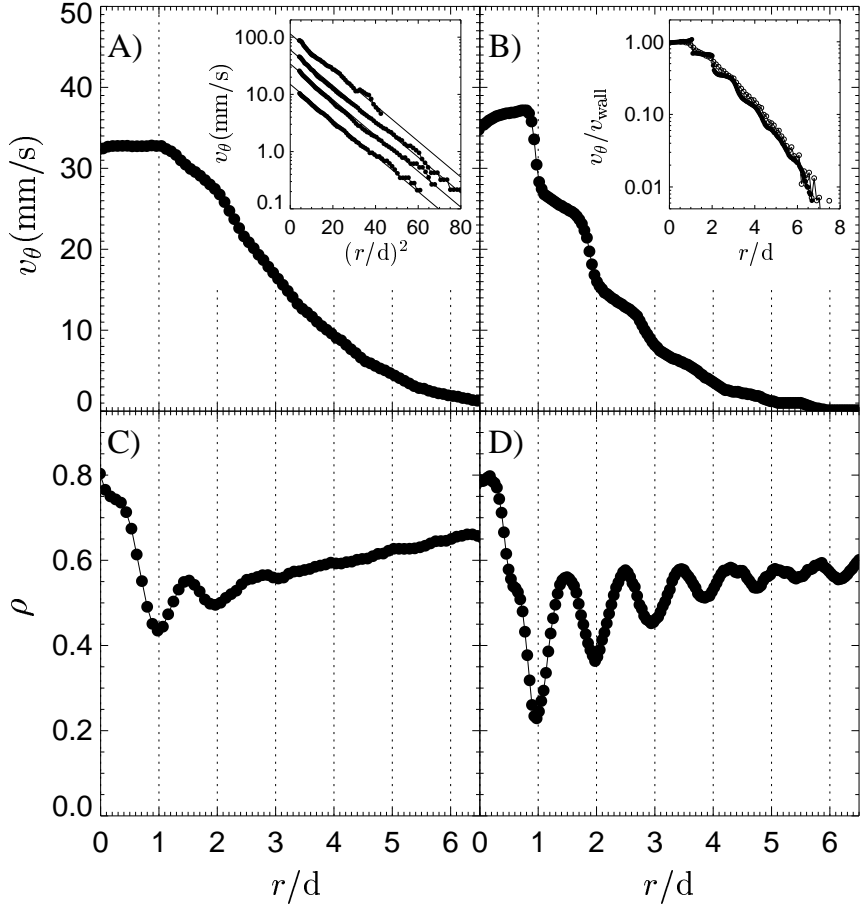,width=4.5in}}
\vspace{2ex}
\caption{ The velocity and density profiles in the bulk for poppy and
mustard seeds adapted from \protect\cite{mueth00}. The radial profile
of the azimuthal velocity $v_\theta(r)$ within the bulk of a 3D sample
of poppy seeds (A) and mustard seeds(B) was previously measured using
MRI. The inner layer of particles of diameter ${\rm d}$, between
$r/{\rm d}=0$ and $r/{\rm d}=1$, was glued to the inner cylinder wall
and thus exhibits a nearly constant velocity profile, while the free
material for $r/{\rm d}>1$ experiences shear flow. The velocity
profile for poppy seeds (${\rm d}=0.8{\rm mm}$) was found to be
Gaussian in shape and rate independent (inset to A). The velocity
profile for mustard seeds (${\rm d}=1.8{\rm mm}$) exhibits sharp drops
and steps because MRI measures the mass flow velocity which differs
from the velocity profile for particle centers if significant layering
of particles occurs.  The velocity profile $v_\theta(r)$ for mustard
seeds in the middle of the cell as determined by MRI was compared to
that on the bottom surface of the cell, as measured by particle
tracking in videos (inset to B).  The comparison reveals that the flow
within the bulk and on the lower surface have the same overall shape.
The density profile inside the 3D system was measured using X-ray
tomography for samples of poppy seeds (C) and mustard seeds (D). While
little ordering is observable for poppy seeds, pronounced oscillations
in $\rho(r)$ for mustard seeds reveals that the particles are layering
along the moving inner cylinder wall. }
\label{fig:poppymustard}
\end{figure}

Several groups have studied the time-averaged steady-state flow
velocity in slow, dense granular Couette flow in both 2D and 3D
~\cite{khosropour97,veje99,howell99,karion00,mueth00,losert00,bocquet01}.
Typically one measures the average particle velocity in the azimuthal
($\hat \theta$) direction, $v_\theta(r)$, as a function of the
distance $r$ from the moving inner cylinder wall.  These studies
reveal that the velocity decays quickly with $r$, with most shear
occurring in a relatively narrow shear band of about 10 grains across
for a broad range of parameters.  High precision MRI measurements in
3D~\cite{mueth00} have shown that $v_\theta(r)$ can be expressed as a
product of an exponential decay and a Gaussian:
\begin{equation}
\label{eqtn:origexpgauss}
v_\theta(r) = v_0 {\rm exp}[ -{\rm b}(r/{\rm d})]{\rm exp}[-{\rm
c}(({r-r_0)/{\rm d}})^2].
\end{equation}
The exponential term was found to be associated with the formation of
layers (see Fig.~\ref{fig:poppymustard}D, adapted from~\cite{mueth00})
which can slip past each other for systems of smooth, monodisperse
spheres. For these materials, the exponential term dominates (see
Fig.~\ref{fig:poppymustard}B) in both 3D~\cite{mueth00,losert00} and
2D~\cite{veje99,howell99} systems.  Note that, although the
exponential term dominates for these systems, the Gaussian term is
still present and gives rise to a characteristic, downward concave
curvature on a log-linear plot (inset to
Fig.~\ref{fig:poppymustard}B).  For systems of particles which are
aspherical, polydisperse, or rough, the exponential term essentially
vanishes and a pure Gaussian velocity profile is observed in
3D~\cite{mueth00}(Fig.~\ref{fig:poppymustard}A). The center of the
Gaussian was found to be at the inner cylinder wall (ie. $r_0 \approx
0$), as seen in the inset to Fig.~\ref{fig:poppymustard}A.  Thus,
Eq.~\ref{eqtn:origexpgauss} can be recast into:
\begin{equation}
\label{eqtn:expgauss}
v_\theta(r) = v_0 {\rm exp}[ -(r/\lambda)-(r/\sigma)^2].
\end{equation}
where the fitting parameters $\lambda$ and $\sigma$ describe the decay
lengths of the exponential and Gaussian components, respectively.
Note that unlike a pure exponential profile, in which $r$ can be
translated to a new origin with a simple rescaling of the velocity,
the Gaussian component has a center position, $r_0 \approx 0$, which
specifies a unique reference location in the system.

In the slow shearing regime, the velocity profile normalized by the
shear rate was found to be invariant to the imposed shear rate at the
inner wall~\cite{howell99,mueth00,losert00}.  Furthermore, using MRI
in 3D the velocity profile $v_\theta(r)$ was found~\cite{mueth00} to
be uniform throughout the height of the cell to within several
particles (the detectable limit) of the upper and lower surfaces.
Comparison of MRI measurements with video data obtained from the lower
cell surface revealed~\cite{mueth00} that the bottom layer of
particles has the same average velocity profile as the interior (see
inset to Fig.~\ref{fig:poppymustard}B).  This is likely caused by the
relative smoothness of the lower boundary surface, contrasted with the
geometrically rough surface of the bulk granular material above the
lowest layer of particles.  Particles on the lower surface are moved
with the bulk material above it since these forces dominate the
relatively small frictional forces on the lower surface.  A direct
comparison of the upper free surface with the bulk was not performed
because significant heaping (several particle diameters) occurred at
the top surface making video tracking of these particles difficult.

Granular Couette flow experiments in 2D and 3D vary in several
important ways.  In 2D, the Couette cell imposes a constant volume and
thus fixed average packing fraction on the system as a whole.  Since
no dilation into the vertical direction (against gravity) can occur,
this leads to a strong dependence of the flow properties on the
average packing fraction~\cite{veje99,howell99}.  In a 3D Couette cell
with free top surface, the packing fraction can adjust freely in
response to gravity, Reynolds dilation, and vibratory compaction until
a local steady-state packing fraction is reached. Thus, the
steady-state behavior of the 3D system does not exhibit the
transitions in behavior observed in 2D systems when the initial
packing fraction is varied.  While both 2D and 3D systems give similar
velocity profiles $v_\theta(r)$ for nearly monodisperse, smooth
spheres, they are found to differ qualitatively for irregularly shaped
particles as experiments in 2D have found a nearly pure exponential
form~\cite{noteirregparticles}.  It has been speculated that this may
be caused by differences in the dimensionality of clusters of
correlated particles~\cite{debregeas00} or possibly by differences in
the way particles rotate (spin) in the two
systems~\cite{howellprivcommun}.

Qualitative properties of the flow can be determined from videos or
from graphs showing the trajectories of particles through time, such
as the one shown in Fig.~\ref{fig:sketch}C, revealing that the motion
is intermittent in time.  These imaging tools also give first hints
that particles are correlated in space: Clusters of particles can be
seen to move together for short intervals.  By watching the motion of
particles which are far from the moving wall, one is able to observe
motion events which involve the collective rearrangement of many
particles in the system, from the inner wall out to larger $r$. This
suggests that buckling of force chains, radiating outward from the
moving wall, causes failure events that result in particle
rearrangements.  Experiments performed in 2D with birefringent disks
have been able to visualize these force chains and their dynamic
evolution directly~\cite{howell99}.

Recent studies by the Haverford/Penn group of the top surface of 3D
Couette flow~\cite{losert00,bocquet01} show that the average flow on
the top surface resembles, at least qualitatively, that observed in
other parts of 3D systems: the velocity profile $v_\theta(r)$ is
between exponential and Gaussian in shape, and for polydisperse
materials $v_\theta(r)$ decays more slowly and is more Gaussian than
for monodisperse materials~\cite{rzeronote}.  These experiments also
measured the RMS velocity fluctuations, $\delta v_\theta(r)$ and
$\delta v_r(r)$, in the azimuthal and radial directions and found that
these fluctuations decayed more slowly with $r$ than does the average
velocity, $v_\theta(r)$.  In addition, the shear rate, $\dot \gamma
\equiv \partial v_\theta / \partial r$, was observed to scale with the
fluctuations as $\delta v_\theta(r) \propto \dot \gamma^\alpha$, where
$\alpha=0.4$.

The upper surface of 3D Couette flow was also studied by the Caltech
group~\cite{karion00}.  They observed a measurable secondary flow
which moved from the outer cylinder to the inner cylinder, and down
the inner cylinder wall, giving rise to heaping on the upper surface.
These experiments confirmed the rate-independence of the azimuthal
flow although no functional form for $v_\theta(r)$ was given.  The RMS
velocity fluctuations about their mean, $\delta v_\theta(r)$ and
$\delta v_r(r)$, were also measured.

Several theories have been put forth to describe granular flow.  The
most well-studied of these is granular kinetic
theory~\cite{jenkins83}, which locally relates the stress to the
fluctuations and energy dissipation using coarse-grained variables.
This theory assumes that collisions are instantaneous and binary,
which typically requires the flow velocity to be large and particle
density to be low.  In this limit, the flow is not expected to be
shear rate independent and $\delta v \sim \dot
\gamma$~\cite{haff86}. In the slow, dense flow limit, on the other
hand, particles often have multiple, persistent contacts and forces
are transmitted along dynamically rearranging force chains. Recently,
Bocquet et al. have proposed~\cite{bocquet01} an extension of kinetic
theory for the dense flow regime.  This continuum model makes some of
the same key assumptions as kinetic theory, but tries to capture the
behavior in the dense limit by proposing a stronger divergence of the
viscosity with packing fraction as the random close packed limit is
approached.  This leads to $\delta v \sim \dot \gamma^\alpha$ with
$\alpha < 1$.

A number of discrete models have been proposed which attempt to
capture the intermittancy and correlation in particle motion seen in
videos of the particle flow and expected for a well-connected system
similar to the static bead packs.  Debregeas and
Josserand~\cite{debregeas00} suggest the material flows in clusters of
various sizes on short timescales and predict that the average flow
profile, $v_\theta(r)$, should be Gaussian in 3D and exponential in
2D.  This is consistent with existing data on irregularly-shaped
particles in 2D~\cite{noteirregparticles} and 3D
systems~\cite{mueth00}, although not for monodisperse, round particles
where layers form and slip past each other.  A second approach
proposed by Josserand~\cite{josserand99}, which also predicts profiles
for $v_\theta(r)$ which vary from exponential to Gaussian, is based on
a lattice model which directly correlates particle motion from one
layer to the next. Here the coupling between particles, possibly
related to surface and geometrical friction, is the control parameter
which determines the form of $v_\theta(r)$. Tkachenko and
Putkaradze~\cite{tkachenko00} model correlated cluster motion in yet
another way, arriving at a form for $v_\theta(r)$ which varies from a
Gaussian to exponential depending on the packing fraction profile to
which the system relaxes.  Each of these models, based on different
physical assumptions, attempts to capture the cooperative
rearrangements observed in videos of the shear flow.

\section*{Experimental Method}

The experimental apparatus used for the studies described in this
article was the same, with only minor modifications, as that used in
previous studies~\cite{mueth00} which measured the radial profile of
the steady-state azimuthal flow, $v_\theta(r)$, at various heights
using MRI.  The granular material was confined between two coaxial
cylinders of diameters 51~mm and 82~mm (see Fig.~\ref{fig:sketch}A).
The outer cylinder and cell floor were held stationary while the inner
cylinder was rotated at constant velocity (the velocity of the inner
wall was $v_{\rm wall} \approx 23 {\rm mm/s}$ for all experiments
described in this article).  To provide a reproducible source of
friction, each cylinder surface had a layer of particles glued to it.
A portion of the lower surface of the cell was imaged (see
Fig.~\ref{fig:sketch}B) at 250~fps and 256x240~pixel resolution using
a monochrome Kodak Motion Corder Analyzer digital camera.

Our previous particle tracking experiments using seeds or glass
spheres showed that the motion was intermittent and correlated in
space (see Fig.~\ref{fig:sketch}C), and that the time-averaged
azimuthal flow profile, $v_\theta(r)$, on the lower surface was
similar to that at other heights within the cell.  However, these
experiments were unable to measure directly the fluctuations in the
particle velocities or the correlations between particle velocities
due to insufficient resolution in determining the exact trajectories
of each particle.

In order to obtain the necessary resolution in measuring the particle
trajectories, a number of modifications were made.  The most important
of these changes was to use polished stainless steel balls
(manufactured for ball bearings).  Illuminating the cell with a single
small light source produced a small, well-defined reflection from each
particle~(see Fig.~\ref{fig:sketch}D) which allowed for a very precise
position measurement.  The high degree of sphericity of ball bearing
balls ($\delta r / r \le 10^{-4}$ for the grade 100 balls used) almost
completely eliminated the false motion of particles observed when less
precise spheres, such as commonly available glass
spheres~\cite{noteglassprecision}, were used. It was also essential to
minimize blemishes in the lower surface which may introduce noise into
particle position measurements.  To maximize the spatial resolution of
each position measurement, I spread out the light reflected from each
particle across as many pixels as possible. This was done by
defocussing the lens, producing a small annulus for each
particle~\cite{noteprecision,thankslosert}.  I determined the
uncertainty $\delta_x$ in each position measurement from all noise
contributions by performing test runs with a system in which all
particles were glued to a disk.  At very slow disc rotation rates, the
noise in position measurement was 0.05~pixels or 4~$\mu{\rm m}$.  By
analyzing video data of completely stationary particles, I found the
same fluctuation in measured position, indicating that this noise
floor in the position measurement was introduced by pixel noise in the
camera.  This was confirmed by directly measuring the pixel noise in
the camera and calculating the uncertainty this introduced into the
center of mass measurement.

To eliminate crystallization, which occurs very rapidly for highly
monodisperse spheres against a flat surface, a bidisperse sample was
used.  The sample consisted of 1.0 and 1.5~mm balls mixed in equal
quantities by weight.  Along the bottom surface and in the primary
shear zone near the moving wall, the larger particles were observed to
gradually move away from the lower surface.  This presumably occurs
because the contact angle between smaller and larger spheres on the
lower surface has a non-zero angle relative to that surface, giving
rise to a vertical component in any contact force. This pushes larger
particles out of the bottom layer of particles.  However, the small
concentration of larger balls which remained on the lower surface,
primarily at larger distances from the moving wall, was sufficient to
prevent crystalline ordering~\cite{nelson82}.  As will be discussed
further below, aside from the lack of crystalline ordering the system
behaved as if it were monodisperse.

Each experimental run consisted of 5459 digital video frames of
resolution 256x240 pixels and imaging approximately 350 particles.
The position of each particle in each frame was measured using a
convolution method which determined the center of brightness of each
reflection~\cite{notegriercrocker}.  From these sets of positions at
each time, the particle trajectories were assembled and the velocities
were determined.  Each experimental run yielded approximately
2,000,000 velocity measurements.  Typically, between three and five
runs of each experimental type were performed. These were either
averaged or compared to guarantee reproducibility of the results.

The average velocity and density profiles and the velocity
correlations were affected very little by random noise. However,
velocity fluctuation measurements do not distinguish between actual
fluctuations in the particle motion and noise in the particle position
measurements.  This noise is caused by various effects including
camera pixel noise, optical artifacts, pixelation, assumptions in
software, and imperfections in the spheres.  To determine the net
contribution of this measurement noise to the fluctuation
measurements, calibration experiments were performed. The calibration
was done by fixing particles to a disk which was moved at constant
angular velocity.  Although the particles move along uniform arcs, the
measured trajectories have a non-negligible noise level which varies
with disk rotation rate.  A wide range of rotation rates was used to
obtain calibrations for use at various locations in the cell, since
the average particle velocity varies through the cell.  These
calibration experiments were analyzed in exactly the same way as the
other experiments.  Using these calibrations, the fluctuation
measurements were corrected, extending the range over which the
velocity fluctuations could be reliably measured significantly.

\section*{Experimental Results}

\subsection*{Average velocity and density profiles}

\begin{figure}
\centerline{
\psfig{file=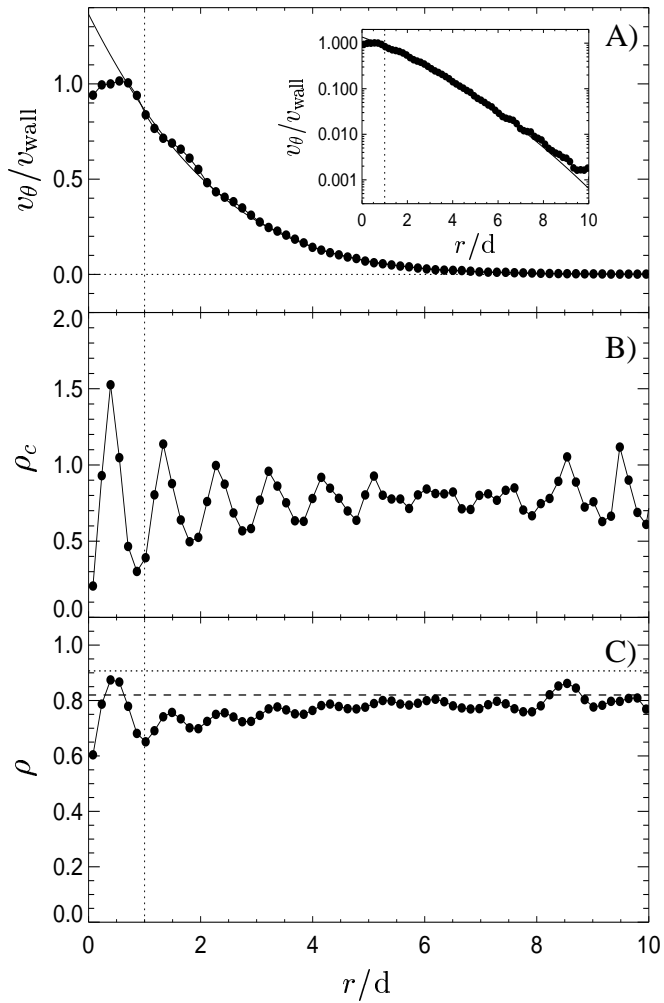,width=3.5in}}
\vspace{2ex}
\caption{ Radial velocity and density profiles. The material between
$r/{\rm d}=0$ and $r/{\rm d}=1$ is glued to the inner wall giving rise
to a nearly constant velocity profile. (A) The velocity profile for
essentially monodisperse bearing balls (${\rm d}=1.0{\rm mm}$), fit to
Eq.~\ref{eqtn:expgauss}. The dashed line at $r/d=1$ represents the
edge of the layer of particles glued to the inner cell wall. When
plotted semi-log (inset), $v(r)$ is seen to be dominated by the
exponential term with very weak curvature, as expected for a nearly
monodisperse sample of smooth spheres. (B) The distribution of
particle centers, $\rho_c(r)$, showing pronounced layering. (C) The
local packing fraction of material, $\rho(r)$, calculated from
$\rho_c(r)$. The horizontal dotted line represents the density of a
crystal in 2D, $\rho_{\rm xtal}^{\rm 2D} = 0.91$, and the horizontal
dashed line represents that of random close packed spheres in 2D,
$\rho_{\rm rcp}^{\rm 2D}=0.82$. }
\label{fig:vrhor}
\end{figure}

The radial profile of the time-averaged azimuthal velocity,
$v_\theta(r)$, was obtained by averaging the velocities of all
particles over all times at each radius $r$ (see
Fig.~\ref{fig:vrhor}A).  I found that $v_\theta(r)$ was well-fit by
Eq.~\ref{eqtn:expgauss} with $\lambda=(2.1 \pm 0.2){\rm d}$ and
$\sigma=(4.7 \pm 0.4){\rm d}$ (solid line in Fig.~3A and inset).  This
profile has a dominant exponential term, similar to measurements of
nearly monodisperse samples of smooth spheres~\cite{mueth00}. The
velocity profile is smooth and does not show the strong steps seen
with MRI experiments, as video techniques track the particle center
motion and not the average flow of material, as MRI does.  For
materials for which the packing fraction profile, $\rho(r)$,
oscillates significantly, such as smooth monodisperse spheres, the
mass flow profile measured by MRI exhibits steps in the velocity
profile (as seen in Fig.~\ref{fig:poppymustard}B).  Note also that,
for $r<1$, the mean flow velocity profile is essentially constant as
these particles are glued to the moving inner cell wall.

The radial distribution of particle center positions can be used to
calculate the particle center packing fraction $\rho_c(r)$
(Fig~\ref{fig:vrhor}B) which is normalized such that $\rho_c(r)=1$
would correspond to all space being occupied by material.  Pronounced
oscillations are visible in $\rho_c(r)$ corresponding to ordering of
the particles into concentric layers.  This layering is generally only
observed for smooth spherical particles and accompanies the
predominantly exponential form of $v_\theta(r)$~\cite{mueth00}.

By convolving $\rho_c(r)$ with a kernel describing the cross-sectional
area of an individual particle, one obtains the material packing
fraction $\rho(r)$ (see Fig.~\ref{fig:vrhor}C).  Because the video
images show the number of particles over a given area, the kernel used
is the cross-section of a disk and both $\rho_c(r)$ and $\rho(r)$ are
2D quantities.  In order to calculate the packing fraction accurately,
the local relative populations of smaller and larger spheres at each
radius $r$ in the cell were measured and used when calculating
$\rho(r)$.  The layering of particles is still observable in $\rho(r)$
although it is much more subtle than in $\rho_c(r)$.  Reynolds
dilation is also visible at small $r$.  At larger $r$, $\rho(r)$
approaches the random close packing fraction in 2D~\cite{bideau86},
$\rho_{\rm rcp}^{\rm 2D}=0.82$. (At the largest $r$, layering occurs
again due to volume exclusion effect near the outer wall.)

\subsection*{Velocity-time traces}

\begin{figure}
\centerline{
\psfig{file=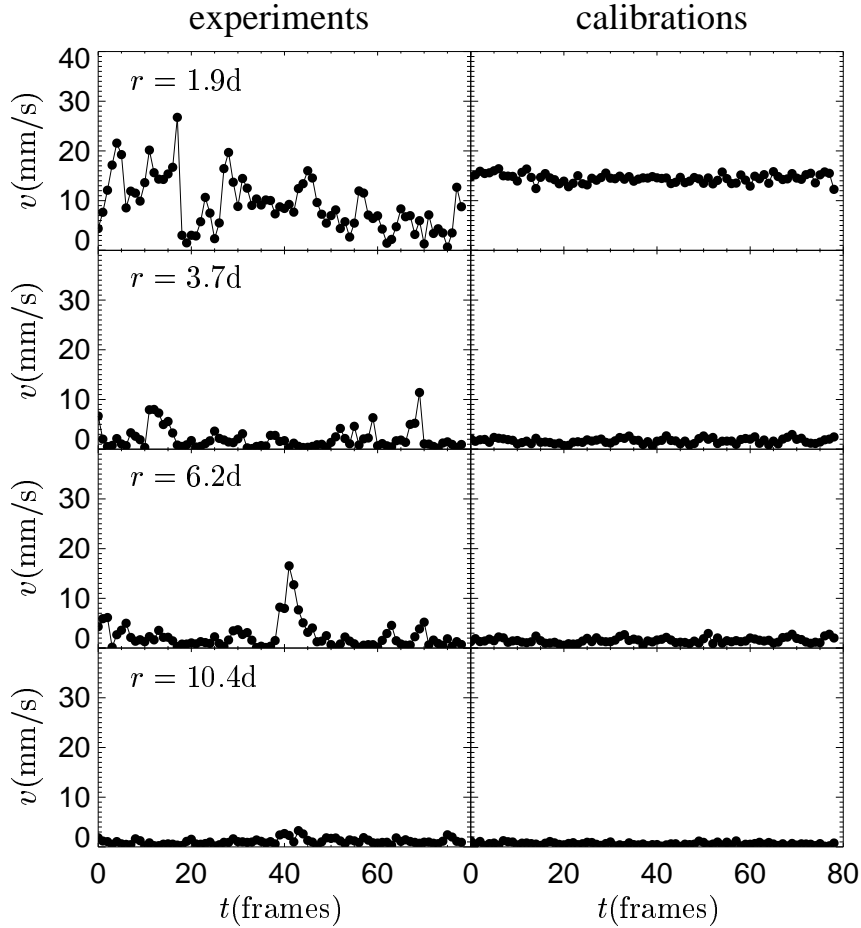,width=4.5in}}
\vspace{2ex}
\caption{ Time traces of the velocity of individual bearing balls at
different radii.  The velocity evolution over time (1 frame is 4~ms)
for four randomly selected particles at radii 1.9d, 3.7d, 6.2d and
10.4d for an experimental run are shown in the left column.  The right
column shows calibration data obtained by tracking particles which
were glued to a disk and rotated at velocities corresponding to the
approximate average velocity for each particle in the experiment (left
column).  The amplitude of the fluctuations in the right column show
the noise level for the given rotation rate.  Distinct ``motion
events'', intervals over which a particle experiences significant
motion, are clearly observable in the experimental runs. The frequency
of the motion events is seen to decrease quickly with distance from
the moving inner wall (at $r=10{\rm d}$ essentially no events occur
during the 80-frame time interval), although the amplitude and
duration only show a small decrease. }
\label{fig:traces}
\end{figure}

A first step toward quantifying the intermittant character of the flow
is to plot the velocity of a single particle over time for randomly
chosen particles at various positions in the cell.  Several of these
velocity traces are shown in Fig.~\ref{fig:traces}.  The left column
shows the velocities of four different particles at different
distances $r$ from the inner wall.  The right column shows calibration
data obtained by rotating a disk with particles glued to it at rates
roughly giving the same average velocity as in the corresponding plot
in the left column.  The calibration data in the right column thus
provides an indication of the velocity-dependent noise level,
observable as the fluctuations in the signal $v(t)$.

The velocity traces for experimental runs (left column) show distinct
motion events during which appreciable particle motion persists.  The
duration of these events is typically several frames (one frame
corresponds to 4~ms), and is roughly independent of radial position in
the system.  While some velocity changes are gradual over several
frames, many large rapid increases and decreases in velocity are
visible.  These presumably correspond to very fast jamming and
unjamming events.  Although their amplitude does not vary strongly
with $r$, the frequency of these events varies substantially,
occurring at increasingly longer intervals for larger $r$.

\subsection*{Velocity distributions}

\begin{figure}
\centerline{
\psfig{file=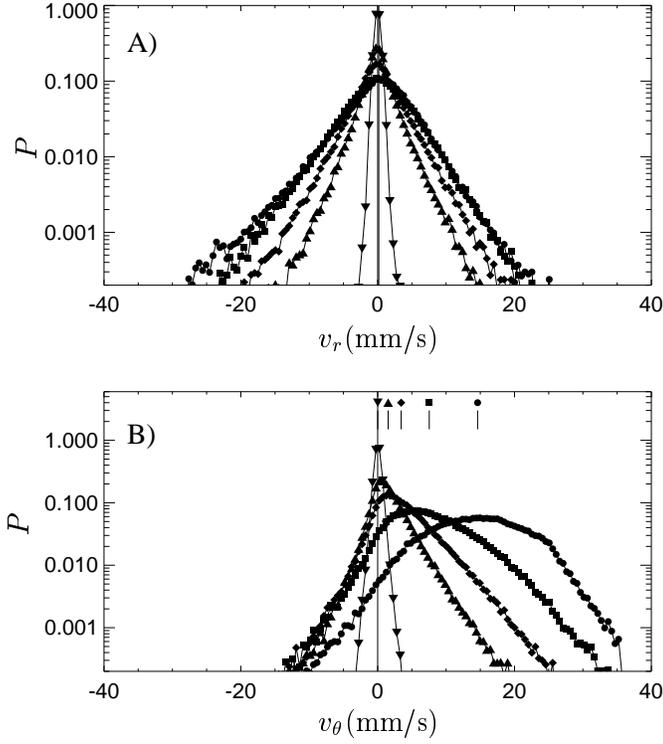,width=3.5in}}
\vspace{2ex}
\caption{ Distribution of particle velocity components. Histograms of
the radial (A) and azimuthal (B) particle velocity components
determined from two successive video frames (ie. averaged over 4~ms)
are shown.  Each curve represent particles at different radii:
$r=1.5{\rm d}$ (circles), $r=2.5{\rm d}$ (squares), $r=3.6{\rm d}$
(diamonds), $r=4.6{\rm d}$ (triangles), $r=9.7{\rm d}$
(down-triangles).  The average velocity in the azimuthal direction is
plotted as small vertical lines at the top of the lower plot, labeled
by the symbol of the curve which it corresponds to. }
\label{fig:veldists}
\end{figure}

The velocity components in the radial direction, $v_r$, and azimuthal
direction, $v_\theta$, were histogrammed for all particles and times
for several ranges of radii $r$ (Fig.~\ref{fig:veldists}).  These
velocities represent the measured average velocity between two
successive video frames (ie. over 4~ms).  The histogram of radial
velocities $P(v_r)$ is peaked at $v_r=0$ and is approximately
symmetric about this point since the primary flow is not in the radial
direction. The width of the distributions decrease with increasing
distance $r$ from the rotating cell wall.  The off-center
distributions $P(v_\theta)$ in (B) reflect the average azimuthal
velocity, $v_\theta(r)$, as seen in Fig.~\ref{fig:vrhor}A.

The average value of $v_\theta$ at each radius is shown by the
position of the corresponding data point at the top of the graph.  The
velocity distributions $P(v_\theta)$ are peaked near, but slightly
below, their average value because there are larger velocity
fluctuations in the forward direction.  Fluctuations away from the
mean velocity decay roughly exponentially, especially far from the
inner cell wall, but acquire a rounded shape for small $r$.

\subsection*{Fluctuation amplitudes}

\begin{figure}
\centerline{
\psfig{file=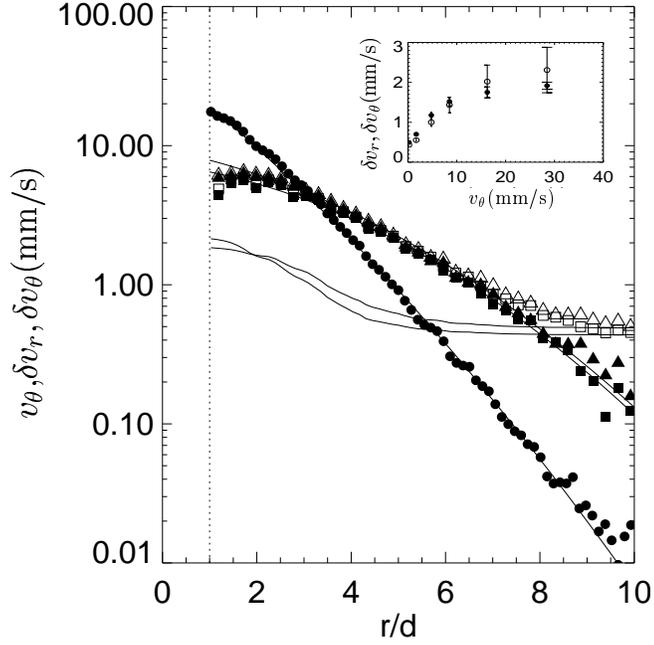,width=3.5in}}
\vspace{2ex}
\caption{ Velocity and velocity fluctuation profiles. The velocity
profile (solid circles) is shown with its fit to
Eq.~\ref{eqtn:expgauss}. The azimuthal and radial RMS velocity
fluctuations about their mean, $\delta v_r(r)$ (open squares) and
$\delta v_\theta(r)$ (open circles) as defined in Eq.~\ref{eqtn:dv}
are also shown. The inset shows the background values of $\delta
v_r(r)$ and $\delta v_\theta(r)$ measured from calibrations performed
by glueing particles to a disk which was rotated at various
speeds. Using this calibration and the average velocity profile
$v_\theta(r)$ (solid circles), the noise floors for $\delta v_r(r)$
and $\delta v_\theta(r)$ were determined (two lower solid curves).
The corrected curves for $\delta v_r(r)$ and $\delta v_\theta(r)$
(solid circles and triangles) were obtained by subtracting the
contribution from the noise floors~\protect\cite{quadrature}. The
corrected fluctuation profiles are shown with fits to
Eq.~\ref{eqtn:dvrexpgauss}.  }
\label{fig:fluctuations}
\end{figure}

The RMS velocity fluctuations in the radial and azimuthal directions
were calculated using the set of all particle velocities collected
over all times.  The velocity fluctuation of the $i$'th particle about
the mean at time $t$ is given by $\overrightarrow{\Delta v_{i,t}}
\equiv \overrightarrow{v_{i,t}}-\overrightarrow{\bar v(r=r_{i,t})}$,
where $\overrightarrow{v_{i,t}}$ is the velocity vector of the $i$'th
particle at time $t$, $r_{i,t}$ is the distance of the $i$'th particle
from the inner wall at time $t$, and $\overrightarrow{\bar v(r)}$ is
the average velocity of all particles at distance $r$ from the inner
wall.  The RMS velocity fluctuations about the mean in the radial and
azimuthal directions, $\delta v_r(r)$ and $\delta v_\theta(r)$, are
given by
\begin{equation}
\label{eqtn:dv}
\delta v_r(r) = \sqrt{ 
{\sum {\Delta {v_r}^2_{i,t}}
\delta(r_{i,t}-r) } 
\over 
{\sum \delta(r_{i,t}-r) } 
}
,
\delta v_\theta(r) = \sqrt{ 
{\sum {\Delta {v_\theta}^2_{i,t}}
\delta(r_{i,t}-r) } 
\over 
{\sum \delta(r_{i,t}-r) } 
}
\end{equation}
where the sums are taken over all particles $i$ and times $t$, and
$\Delta {v_\theta}$ and $\Delta v_r$ are the azimuthal and radial
components of $\overrightarrow {\Delta v}$.  The RMS fluctuation
amplitudes $\delta v_r(r)$ (open squares) and $\delta v_\theta(r)$
(open circles) are shown in Fig.~\ref{fig:fluctuations}, along with
the average azimuthal flow $v_\theta(r)$ (solid circles).  The average
fluctuation amplitude is seen to be somewhat smaller than
$v_\theta(r)$ in the primary shear zone, but quickly dominates the
azimuthal flow for $r > 3$.

By measuring the RMS fluctuation amplitude in calibration experiments
taken at varying rotation rates (see inset to
Fig.~\ref{fig:fluctuations}), the experimental noise floor for $\delta
v_r(r)$ and $\delta v_\theta(r)$ was obtained.  The noise floors for
$\delta v_r(r)$ and $\delta v_\theta(r)$ are shown by the two lower
solid curves in Fig.~\ref{fig:fluctuations}.  By removing the
contribution from the measurement noise from these curves, the
corrected particle velocity fluctuation amplitudes (solid triangles
and squares in Fig.~\ref{fig:fluctuations}) were
obtained. Empirically, the corrected fluctuation data is well-fit by
\begin{equation}
\label{eqtn:dvrexpgauss}
\delta v_\alpha(r) =
\delta v_{\alpha,0}{\rm exp}[-(r/\lambda_\alpha)-(r/\sigma_\alpha)^2]
\end{equation}
where $\alpha \in \{r,\theta\}$. The fitting parameters for the radial
fluctuations $\delta v_r(r)$ were found to be $\delta v_{r,0} =
(7.4\pm 0.6) {\rm mm/s}$, $\lambda_r = (10 \pm 0.9){\rm d}$, and
$\sigma_r = (5.8 \pm 0.5) {\rm d}$, while the parameters for $\delta
v_\theta(r)$ were $\delta v_{\theta,0} = (9.0 \pm 0.8){\rm mm/s}$,
$\lambda_\theta = (6.7 \pm 0.5){\rm d}$, and $\sigma_\theta = (5.8\pm
0.5){\rm d}$.

\subsection*{Spatial correlations}

To identify quantitatively whether there are correlations, and to
determine the lengthscale over which these correlations exist, the
spatial correlation function between particles at a given radius $r$
was calculated.  The spatial correlation function $C_s(s,r)$ is the
average correlation of the fluctuation of the velocity components
about their means of two particles at the same radius $r$ with a
separation $s$ along an arc of constant radius:
\begin{equation}
\label{eqtn:Cs}
C_s(s,r)={
{\sum
(\Delta {v_\theta}_{i,t} \Delta {v_\theta}_{j,t}
 + \Delta {v_r}_{i,t} \Delta {v_r}_{j,t} )
\delta(\vert r(\theta_{i,t}-\theta_{j,t}) \vert -s) \delta(r_{i,t}-r)
\delta(r_{j,t}-r)}
\over{\sum \delta(\vert r(\theta_{i,t}-\theta_{j,t}) \vert -s) \delta(r_{i,t}-r)
\delta(r_{j,t}-r)}
}
- ({\delta v(r)}_{\rm cal})^2 \delta(s)
\end{equation}
where the sum is over all particle pairs ($i$,$j$) and times $t$.  The
term $(\Delta {v_\theta}_{i,t} \Delta {v_\theta}_{j,t} + \Delta
{v_r}_{i,t} \Delta {v_r}_{j,t} )$ is analogous to a dot product,
however it is performed in polar coordinates so that it asymptotes to a
constant at large particle separations $s$.  The term $\delta
v(r)_{\rm cal}$ is the measurement noise at a given radius (determined
by the average velocity at this radius) as determined from the
calibration experiments.  The final term, $({\delta v(r)}_{\rm cal})^2
\delta(s)$, is subtracted because noise is correlated for
$s=0$~\cite{csdetail1}.  Here I have made the assumption that the
noise of two different particles is uncorrelated.

\begin{figure}
\centerline{
\psfig{file=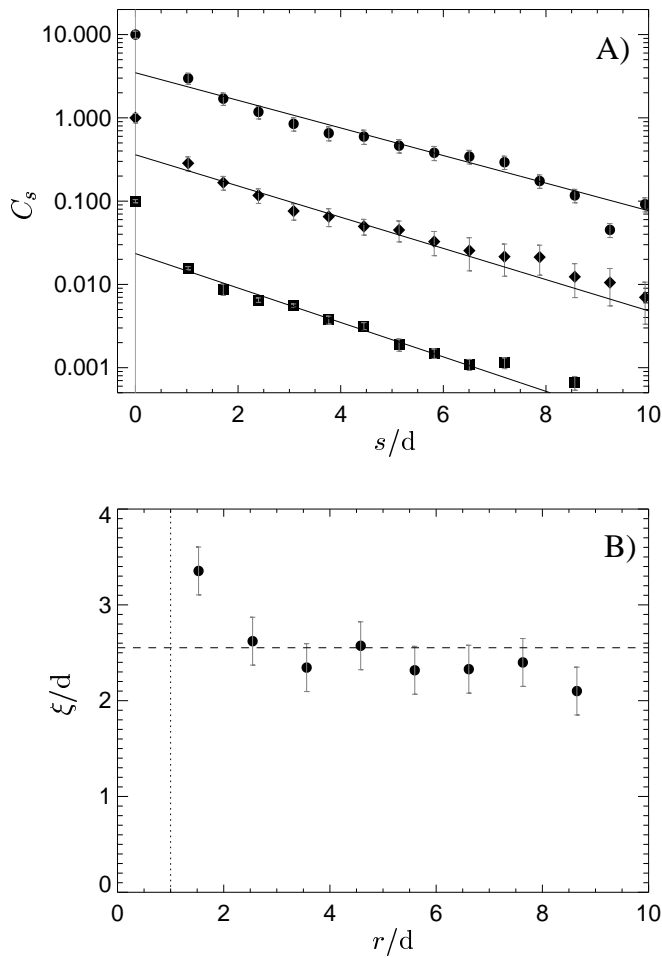,width=3.5in}}
\vspace{2ex}
\caption{ Spatial correlations and correlation lengths. (A) The
correlation $C_s(s,r)$ of the velocity fluctuations of two particles
(defined in Eq.~\ref{eqtn:Cs} in the text) at the same radius $r$ as a
function of their separation $s$ for particles at distances from the
inner wall of 2.5d (circles), 5.6d (diamonds), and 8.6d (squares). The
curves were normalized such that $C_s(s=0)=1$, however they have been
shifted vertically for clarity.  Fits to exponentials for the data
$s/d>0$ are shown, and the slopes are used to calculate the
correlation length $\xi/d$ as shown in (B). (Note that, for clarity,
not all curves were shown in (A).) The correlation length is seen to
be nearly constant outside the primary shear zone and somewhat larger
very close to the inner wall. The average of the data for $r/{\rm
d}>3$ is shown as the horizontal dashed line. }
\label{fig:spacecor}
\end{figure}

We find that $C_s(s,r)$ decays exponentially with particle separation
$s$ for each radius $r$ (Fig.~\ref{fig:spacecor}A).  The deviation
from the exponential at $s=0$ is likely due to motion of particles
which are not in solid contact with neighboring particles, giving an
``extra'' contribution to $s=0$ only. The overall exponential form
suggests that correlations are dominated by nearest neighbor
interactions.

The slopes of the curves in Fig.~\ref{fig:spacecor}A give the
correlation length at each radius, $\xi(r)$, as shown in
Fig.~\ref{fig:spacecor}B.  This correlation length is found to be a
constant, $\xi = (2.2 \pm 0.1) {\rm d}$, outside the primary shear zone.
Within the primary shear zone $\xi(r)$ is seen to increase to a value
of nearly $3.5 {\rm d}$ for the innermost particles. This indicates
that particle move in larger clusters near the inner wall than at
larger $r$.

\subsection*{Time correlations}

In addition to spatial correlations, temporal correlations are
indirectly visible in videos as well as in particle velocity traces
such as those shown in Fig.~\ref{fig:traces}.  To extract the
characteristic timescale, and spatial dependence of these
correlations, I calculated the particle velocity autocorrelation
function, $C_t(\Delta t,r)$, given by the average correlation of a
particle's velocity fluctuation about its mean at times $t$ and
$t+\Delta t$:
\begin{equation}
\label{eqtn:Ct}
C_t(\Delta t,r)={
{\sum
(\Delta {v_\theta}_{i,t} \Delta {v_\theta}_{i,t+\Delta t}
 + \Delta {v_r}_{i,t} \Delta {v_r}_{i,t+\Delta t} )
\delta({1 \over 2}(r_{i,t}+r_{i,t+\Delta t}) -r)
}
\over{\sum
\delta({1 \over 2}(r_{i,t}+r_{i,t+\Delta t}) -r)
}
}
- ({\delta v(r)}_{\rm cal})^2 \delta(\Delta t).
\end{equation}
The sum is taken over all all particles $i$ and times $t$. As with
$C_s(s,r)$, the last term is added to eliminate the contribution of
noise at $\Delta t=0$.  Here I make the assumption that the noise in
measuring a particle's velocity at two different times is
uncorrelated.

\begin{figure}
\centerline{
\psfig{file=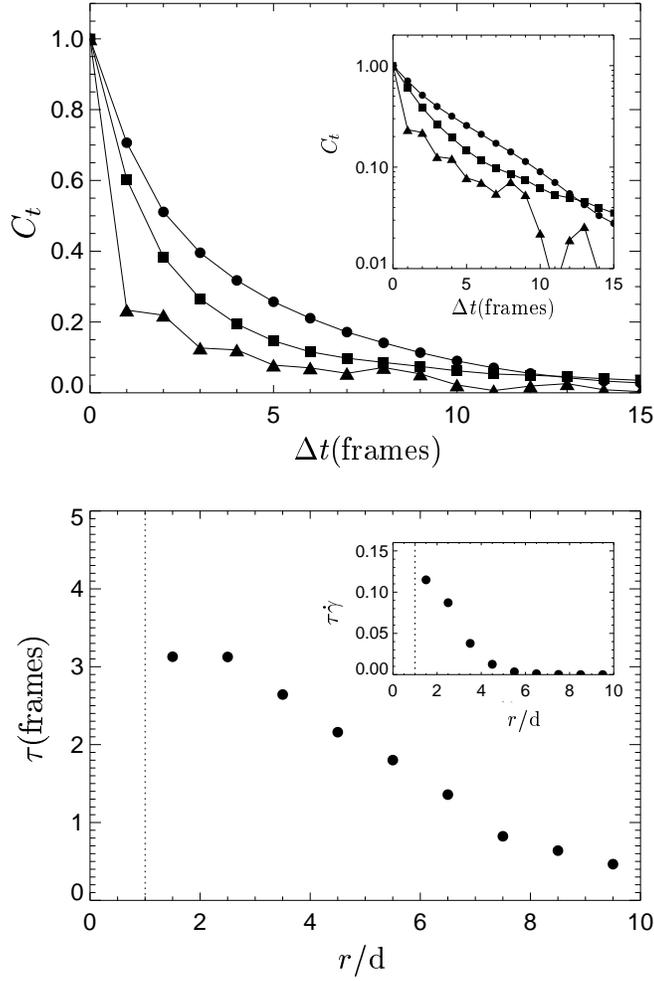,width=3.5in}}
\vspace{2ex}
\caption{ The autocorrelation of a particle's velocity fluctuations
over time.  The top plot shows the autocorrelation function,
$C_t(\Delta t,r)$, of a particle's velocity fluctuations about its
mean at times $t$ and $t+\delta t$ for $r/{\rm d}=1.5$ (circles),
$r/{\rm d}=4.5$ (squares), and $r/{\rm d}=7.5$ (triangles) as defined
in Eq.~\ref{eqtn:Ct}. The noise contribution for $\Delta t=0$ has been
subtracted and $C_t(\Delta t,r)$ is normalized (for each $r$ value)
such that $C_t(\Delta t=0,r)=1$. The correlation time $\tau(r)$,
defined as the time it takes for $C_t(\Delta t,r)$ to fall to $1/e$
for each $r$, is shown in the bottom plot.  The inset shows the
dimensionless quantity $\tau \dot \gamma(r)$, which represents the
duration over which particle motion is correlated relative to the
timescale over which shear occurs.}
\label{fig:timecor}
\end{figure}

$C_t(\Delta t,r)$ is shown in Fig.~\ref{fig:timecor}A for several
distances $r$ from the inner wall. The rate of decay of $C_t(\Delta
t,r)$ is seen to increase with distance $r$ from the inner wall.  The
shape of this correlation function is non-exponential (inset to
Fig.~\ref{fig:timecor}A), indicating that there are correlated
processes occurring at multiple timescales. Nevertheless, we can
define an effective timescale $\tau(r)$ as the time it takes for
$C_t(\Delta t,r)$ to fall to ${1 \over e}C_t(\Delta t=0,r)$.
$\tau(r)$ decreases substantially with increasing distance $r$ from
the moving wall (Fig.~\ref{fig:timecor}B).  Thus, particles near the
inner wall typically retain their velocity history for several video
frames while particles far from the inner wall rarely retain their
velocity history for even the interval between two successive frames.
The inset to Fig.~\ref{fig:timecor}B shows the dimensionless quantity
$\tau\dot\gamma(r)$, corresponding to the ratio of the correlation
time scale and the shear flow time scale.

\subsection*{Fluctuation - shear rate relationship}

\begin{figure}
\centerline{
\psfig{file=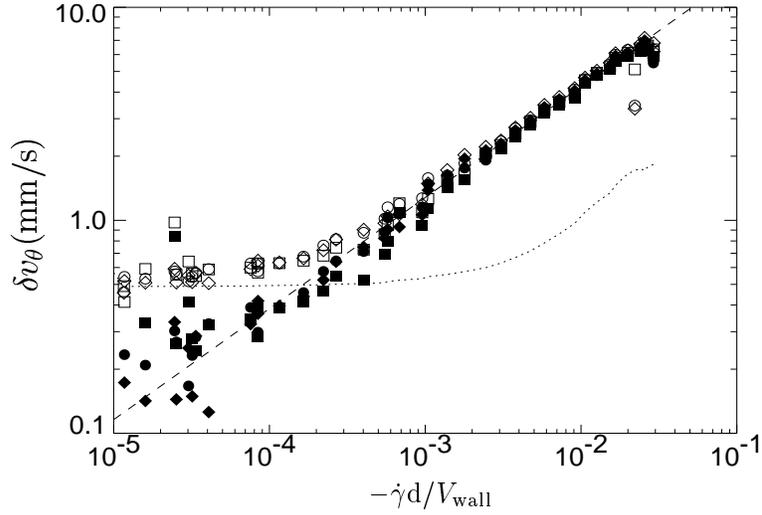,width=4.0in}}
\vspace{2ex}
\caption{ Relationship between the fluctuations in the azimuthal
velocity and shear rate. The RMS fluctuations of the azimuthal
velocity component about its mean, $\delta v_\theta(r)$, are plotted
against the shear rate $\dot \gamma(r) \equiv \partial v_\theta(r) /
\partial r$ for three different experiments (open symbols).  The noise
floor for $\delta v_\theta$ is shown by the dotted curve, revealing
that $\delta v_\theta$ is saturated by the noise floor at low shear
rates.  After removing the contribution of the noise floor to $\delta
v_\theta$~\protect\cite{quadrature} (solid symbols), I found that the
data is well-fit by a power law (dashed line), with a slope of $\alpha
= 0.52 \pm 0.04$. }
\label{fig:shearfluct}
\end{figure}

Having measurements of both the average azimuthal velocity
$v_\theta(r)$ and the fluctuations of the azimuthal velocity about the
mean $\delta v_\theta(r)$, I was able to determine the relationship
between the fluctuation strength and the shear rate. The shear rate is
given by $\dot\gamma \equiv \partial v_\theta(r) / \partial r$.  The
fluctuation - shear rate relationship is shown in
Fig.~\ref{fig:shearfluct}.  The open symbols show the uncorrected
data, which exhibit a power-law-like relationship over a small region
for large shear, and which asymptote to a constant for smaller shear.
The noise floor for $\delta v_\theta(\dot\gamma)$ is calculated from
the calibration curves $\delta v_\theta(v_\theta)$ as shown in the
inset to Fig.~\ref{fig:fluctuations}. This noise floor is shown by
the dotted line in Fig.~\ref{fig:shearfluct}.  For smaller shearing
rates, the fluctuations $\delta v_\theta$ are dominated by the
measurement noise.  After the noise contribution was
subtracted~\cite{quadrature} (solid symbols), the overall power law
shape of the curve is seen to extend into the smaller shearing region
(ie. larger $r$).  The dashed line shows a fit of the
high-shearing-rate portion of the curve to a power law, with a
measured slope $\alpha = 0.52 \pm 0.04$.  This fit is consistent with
the entire range of data.

\subsection*{Shear rate - density relationship}

\begin{figure}
\centerline{
\psfig{file=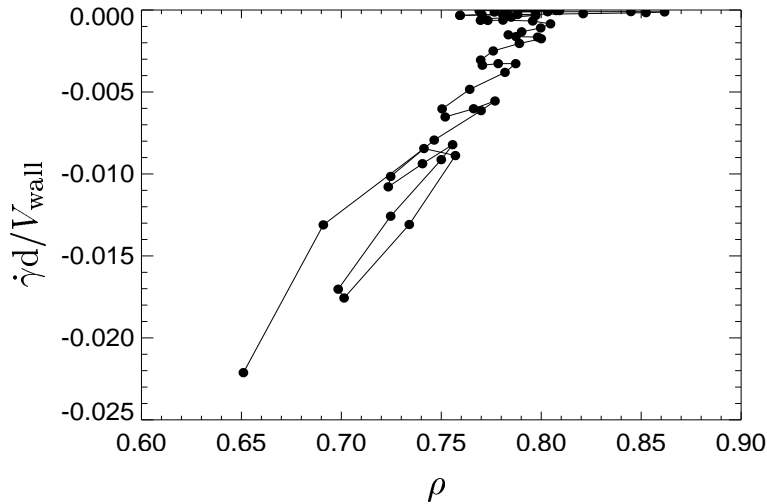,width=4.0in}}
\vspace{2ex}
\caption{ Relationship between the shear rate and packing fraction.
The shear rate $\dot \gamma {\rm d}/V_{\rm wall}$ is plotted against
the packing fraction $\rho$. $V_{\rm wall}=23{\rm mm/s}$ as with all
video tracking experiments described here. Points connected by line
segments correspond to data points at neighboring values of $r$, which
parameterizes both quantities. The lower left data point corresponds
to smallest $r$, while large $r$ values have shear rates approaching
0. For larger values of $r$, the shear rate is nearly 0 on this linear
plot, however the density continues to oscillate substantially. Note
that the three datapoints with largest $\rho$ correspond to the data
in Fig.~3 near the outer wall that exceeds the random close pack
density in 2D, and are likely caused by layering near the outer
cylinder wall.}
\label{fig:gammarho}
\end{figure}

Using the measurements of the material packing fraction, $\rho(r)$,
and material velocity, $v(r)$, for this system, the relationship
between the shearing rate, $\dot \gamma(r)$, and packing fraction,
$\rho(r)$, was readily obtained (Fig.~\ref{fig:gammarho}).  The curve
is parametrized by $r$, with the smallest $r$ value having low density
and high (negative) shearing rate.  The line segments connecting data
points reflects the path followed as $r$ is increased.  Both $\rho$
and $\dot \gamma$ oscillate with increaseing $r$.  Note that while it
is not visible on this figure, $\dot \gamma$ is varying by more than
three orders of magnitude.  For $r$ above 6 particle diameters, $\dot
\gamma$ is nearly zero on this figure although $\rho$ is still
oscillating significantly. While one may approximate this as a linear
profile, there are strong oscillations on the single-grain lengthscale
throughout the cell.

\section*{Discussion}

\subsection*{Average flow properties}

The average velocity, $v_\theta(r)$, was found to be very similar to
that previously obtained~\cite{mueth00} for roughly monodisperse
mustard seeds in the same cell. For all data, Eq.~\ref{eqtn:expgauss}
provided an excellent fit. The nearly monodisperse bearing balls were
best fit using $\lambda = (2.1 \pm 0.2){\rm d}$ and $\sigma = (4.7 \pm
0.4){\rm d}$ while the mustard seeds were fit by $\lambda = (2.5 \pm
0.2){\rm d}$ and $\sigma = (4.1 \pm 0.3){\rm d}$.  The similarity in
the fitting parameters for these two experiments supports the previous
observation that the lower surface has a similar flow profile to the
bulk and indicates that the small polydispersity of the mustard seeds
does not have a dramatic effect on the overall flow behavior.
Although it was not measured in this set of experiments, I expect that
this form of the velocity profile is independent of shearing rate for
low shearing rates and extends into the bulk of the material, as was
previously observed with mustard seeds.

This velocity profile is qualitatively similar to that found by the
Duke group for slightly bidisperse disks in 2D~\cite{veje99,howell99}
and by the Haverford/Penn group for the top surface of a 3D system of
glass spheres~\cite{bocquet01}.  Although a fitting form for
$v_\theta(r)$ was not given for the Haverford/Penn data for glass
spheres, inspection of Fig.~5 in \cite{bocquet01} reveals that the
overall shape of $v_\theta(r)$ exhibits curvature (on a log-linear
plot) similar to what we observe in the inset to Figs.~2b and 3a.
However, the overall velocity gradient across the shear band is
dramatically smaller: $v_\theta(r)$ in Fig.~\ref{fig:vrhor} changes a
factor of 1000 between the inner wall and $r=10{\rm d}$ while in the
experiment by Bocquet et al. it drops by only a factor of 50 over the
same interval. It is possible that this difference is attributable to
the difference between the top surface and the bulk.  I note, however,
that experiments on steel spheres by Bocquet et al. have a rate of
decay similar to my experiments on steel spheres.  Thus, it is also
possible that this difference is attributable to the large
polydispersity (diameters range from 0.55 to 0.95~mm) of the glass
spheres used by Bocquet et al.  In the 2D experiments by the Duke
group, the velocity $v_\theta(r)$ decays much more quickly than in our
experiments.  It is not surprising that the 2D experiments differ
because they are under a constant volume boundary condition which
causes the overall behavior of the system to vary strongly with
packing fraction~\cite{veje99,howell99b}.  The steep packing fraction
gradient at low $r$ is likely to give rise to the steep velocity
gradient.

A large degree of particle layering is observed in the distribution of
particle center positions $\rho_c(r)$ (Fig.~\ref{fig:vrhor}B). As
discussed in~\cite{mueth00}, this layering is the source of the
exponential term in the velocity profile (Eq.~\ref{eqtn:expgauss}).
The calculated material packing fraction $\rho(r)$ in
Fig.~\ref{fig:vrhor}C is a 2D quantity in that it is the number of
particles per unit area along the lower surface of the cell.  This
makes quantitative comparison with the previously measured packing
fraction for mustard seeds (reproduced in
Fig.~\ref{fig:poppymustard}D) inside the bulk difficult. However, the
two are qualitatively very similar.  Oscillations in $\rho(r)$ are
seen to correspond to layering of the particles.  The average packing
fraction gradually increases with distance from the inner wall. The
overall form of $\rho(r)$ appears to be approaching $\rho_{\rm
rcp}^{\rm 2D}$.  The oscillations in the packing fraction are highly
significant, with $\rho_{\rm rcp}^{\rm 2D} - \rho(r)$ typically
varying by a factor of 2 over any interval the size of a single grain.

Although it is widely accepted that the packing fraction of a granular
material determines, in part, the ease at which it can shear, the
relationship between the local shearing rate, $\dot \gamma(r)$, and
local packing fraction, $\rho(r)$, had not been previously measured to
my knowledge.  The observed relationship between these two quantities
(Fig.~\ref{fig:gammarho}) reveal that the overall shape is roughly
linear for small $r$.  This is consistent with the predictions of
Bocquet et al.~\cite{bocquet01}.  Large oscillations in $\rho(r)$ and
$\dot \gamma$ give rise to a substantial degree of variation from the
linear form, however. This emphasizes that these quantities are
varying significantly on sub-particle lengthscales and suggests that
coarse-graining of these quantities may not capture the full behavior
of the system.

\subsection*{Fluctuations}

The RMS fluctuations of the velocity components about their mean were
found to show significant deviation from pure exponential
behavior. Instead, $\delta v(r)$ was well-fit by a Gaussian centered
on $r=0$ (Eq.~\ref{eqtn:expgauss}).  The smooth downward curvature
seen in the fluctuation profiles, $\delta v_r(r)$ and $\delta
v_\theta(r)$, appears to be in contrast to the form proposed by the
Haverford/Penn group~\cite{bocquet01}.  They predict fluctuation
profiles that are constant for $r<r_w$, and fall off as $\delta v \sim
{\rm cosh}({H-y \over {\rm const}})$ for $r>r_w$, where $r_w$ is a
fitting parameter. This form has an elbow at $r_w$ and a concave up
shape for $r>r_w$, in contrast to our observations.  Careful
observation of Fig.~5 in ref.~\cite{bocquet01} suggests that the
fluctuation profiles found in the Bocquet experiments on glass spheres
may also be concave down.

The relationship between the fluctuation strength, $\delta
v_\theta(r)$, and the shear rate, $\dot \gamma(r)$, is consistent with
a power-law (see Fig.~\ref{fig:shearfluct}): $\delta v_\theta \propto
\dot\gamma^\alpha$ with $\alpha = 0.52 \pm 0.04$.  Since the
fluctuation strengths are nearly identical in the radial and
azimuthal directions (as shown in Fig.~\ref{fig:fluctuations}), this
can be expressed in terms of an effective granular temperature, $T
\equiv m {(\delta v)}^2$, obtaining $T \propto \dot\gamma^{2\alpha}$.
With the measured value for $\alpha$, these results are consistent
with $T \propto \dot\gamma$.
  
Several theories make predictions about the shear-fluctuation
relationship with which these results can be compared.  Granular
kinetic theory for fast, dilute flows predicts that $\dot \gamma
\propto \delta v \propto \sqrt T$ for steady shear
flow~\cite{haff86b}.  This is inconsistent with our observations for
slow, dense flows.  This inconsistency is not surprising, as several
of the assumptions of kinetic theory do not hold for slow, dense
flows.  The modified hydrodynamic model of Bocquet et al. predicts
$\delta v \propto \dot\gamma^\alpha$, although it makes no prediction
for the value of $\alpha$ other than $\alpha<1$. They measure $\alpha
= 0.4$ and point out that the 2D experiments of the Duke group have
$\alpha=0.5$. Thus, the observed relationship between the shear rate
and fluctuation amplitude is in agreement with these other experiments
and the modified hydrodynamic model.  From this it is clear that
kinetic theory is unable to capture the behavior in the slow, dense
limit unless the pair correlation function $g(r)$ (and thus the
viscosity) for a dense material is used, as done by Bocquet et al.

\subsection*{Correlations}

The coherent motion of neighboring particles observable in videos of
the motion and trajectory plots, is quantified by the spatial
correlation function $C_s(s,r)$ (Fig.~\ref{fig:spacecor}).  The
exponential decay of $C_s(s,r)$ suggests a process due to nearest
neighbor interactions of particles, such as a string of particles in
contact, correlated over a typical length $\xi$.

The correlation length $\xi(r)$ was found to be essentially constant
away from the inner wall, suggesting the local material property, as
it pertains to coherent motion of particles, is uniform in this
region.  One may speculate that the correlation length $\xi$ may be
determined by the randomness of particle contact angles. Knowledge of
the direction of motion of a given particle is quickly lost by its
neighbors in a disordered pack. In this picture, the constant value of
$\xi(r)$ would correspond to a uniformly disordered pack.  Only inside
the first moving layers, closest to the wall, the correlation length
is found to increase indicating that particles are moving in larger
clusters.  This may be a consequence of particle layering.  The
particles tend to line up with contacts in the same direction as their
primary motion, possibly leading to an increased correlation length.


The coherence of particle motion in time was also originally evident
in videos of the shear flow and particle trajectories. Plots of the
velocity traces for individual particles (Fig.~\ref{fig:traces})
reveal the intermittancy of the flow.  Distinct motion events are
visible at all regions of the cell. In many cases, sudden changes in
velocity indicate that a particle suddenly became jammed or freed.
While the interval between motion events was observed to increase
substantially with distance $r$ from the inner wall, the duration and
speed of each motion event appeared to decrease only slightly with
$r$.

The correlation of a particle's velocity fluctuations over time,
$C_t(\Delta t,r)$ decays with a correlation time $\tau(r)$ which
varies with $r$.  For small $r$, the velocity persists for several
time frames (about 15~ms) for an inner cell wall speed of $v_{\rm
wall} = 23{\rm mm/s}$.  This corresponds to a correlation over the
time it takes the inner wall to move by 0.35d.  Note that while the
correlation time $\tau(r)$ is decreasing with $r$, the characteristic
timescale for shear flow, $\dot \gamma^{-1}$, is increasing with $r$.
Thus, ratio of the correlation timescale to the shear timescale, $\tau
\dot \gamma(r)$, decays very quickly with $r$ indicating that the flow
is not uniformly slowing down with increasing $r$.  The decrease in
correlation time $\tau(r)$ with $r$ may be caused by multiple effects
including the speeding-up of motion events, the weakening magnitude of
the events, and the increased rarity of motion events.  However, it is
clear from looking at velocity-time traces such as those shown in
Fig.~4 that the primary change in behavior is that the interval
between motion events becomes very large at large $r$.  While one may
expect the correlation time $\tau(r)$ to increase with $r$ since the
flow is occuring over longer timescales, the opposite trend is
observed.  With increasing $r$, the events occur increasingly quickly
and at increasingly long intervals.

\section*{Summary and Conclusion}

I found that fluctuations in the velocities of two particles at radius
$r$ are correlated within a characteristic lengthscale $\xi(r)$. Away
from the moving inner wall, the correlation length is $\xi = (2.3 \pm
0.1){\rm d}$, while near the moving wall the correlation length
increases to $\xi \approx 3.5{\rm d}$.  Correlations in time are also
observed, both through velocity traces as well as direct calculation
of the correlation function, $C_t(\Delta t,r)$.  The correlation time,
$\tau$, was found to be 13~ms near the moving inner wall and decreases
with $r$. Distinct motion events, intervals over which a previously
stationary or slow-moving particle moves more quickly before slowing
down or coming to rest, were also directly observed in velocity
traces.  The interval between motion events was observed to increase
quickly with distance from the inner wall.  The observed intermittancy
and velocity correlations are reminiscent of that observed in force
chains in a similar geometry by Behringer and Howell~\cite{howell99},
which suggests force chains may play a role in the dynamics.

Having directly measured the velocity, velocity fluctuations, and
density within the same system, I was able to compare my observations
with the predictions of hydrodynamic and kinetic theories of granular
flow.  I measured a power law relationship between shear rate and
fluctuation amplitude, $\delta v \propto \dot\gamma^\alpha$, with
$\alpha = 0.52 \pm 0.04$.  This form is consistent with the
hydrodynamic theory of Bocquet et al. and experiments by this
group~\cite{bocquet01}, although it is inconsistent with simple
kinetic theory without corrections for high density packs which
predicts $\alpha=1.0$~\cite{haff86b}.  The overall form for the
azimuthal velocity fluctuations about their mean, $\delta
v_\theta(r)$, was found to be Gaussian in shape, in contrast to the
cosh form predicted by the hydrodynamic theory.

The relationship between the shear rate and packing fraction can be
approximated by a linear relationship close to the shearing
wall. However, the variations from this form are quite large due to
the strong oscillations in both the packing fraction and the shear
rate with distance from the inner cell wall.  This emphasizes that the
behavior on sub-particle lengthscales is very different from that on
larger scales, and that the coarse-grained approach used in kinetic
and hydrodynamic theories is unable to describe the behavior on this
scale.

While some aspects of the flow are described within existing theories,
many of the observations reported are still not understood.  The
Gaussian component to the velocity profile, $v_\theta(r)$, indicates
that the inner cylinder wall, $r=0$, plays an important role in the
flow behavior.  As $r$ increases from this wall, the shape of the
azimuthal velocity distributions, $P(v_\theta)$, varies from roughly
Gaussian to exponential in shape (Fig.~5).  Also, although the time
for flow to occur becomes very large at large $r$, the correlation
time $\tau(r)$ decreases with $r$ and changes only by a factor of six
(Fig.~8), as the interval between motion events becomes very large but
the duration of motion events does not change dramatically (Fig.~4).
Together, these results suggest the presence of an underlying
mechanism of shear transmission from the inner wall ($r=0$) outward in
a stochastic fashion along intermittent chains of contact.

\section*{Acknowledgements}
I thank Heinrich Jaeger and Sidney Nagel for their invaluable insight
and support. I also thank Georges Debregeas, Sue Coppersmith, Bruce
Winstein, Christophe Josserand, Vachtung Putkaradze, Alexii Tkachenko,
Dan Blair, Wolfgang Losert, and Jerry Gollub for illuminating
discussions.  This work was supported by the National Science
Foundation under Grant No. CTS-9710991 and by the MRSEC Program of the
NSF under Grant No. DMR-9808595.


\bibliography{couette_tracking,granulates}

\begin{thebibliography}{10}
\bibitem{noteirregparticles}
In 2D Couette systems, sheared under constant volume conditions, rough
  particles such as pentagons may actually produce a velocity profile that is
  very well fit by a pure exponential. (Dan Howell and Bob Behringer, private
  communication).
\bibitem{noteglassprecision}
Commonly available glass spheres are molded and typically have sphericities of
  about 95\%. Gluing these glass spheres to a plate and rotating the plate at
  constant velocity shows an apparent relative motion of the particles easily
  visible to the eye as if the particles were not actually glued in place. This
  increases the uncertainty in each velocity measurement substantially, raising
  the noise floor for the fluctuation measurements, and increasing the noise
  level in the correlation measurements substantially.
\bibitem{noteprecision}
The maximum achievable precision for an annulus of radius $r_a$ pixels is
  approximately $0.25 \over \sqrt{\pi}r_a$ pixels. The realizable precision
  however may be limited by the quality of the optics, sphericity of the
  particles, pixel noise in the camera, and other factors.
\bibitem{thankslosert}
I would like to thank Wolfgang Losert who pointed out the benefits achieved
  from defocussing the optics.
\bibitem{csdetail1}
Correlations between particle velocities at seperate times or positions are
  robust against noise, since the noise is uncorrelated. However for $s=0$,
  velocity measurement noise is additive in $C_s(s,r)$. To eliminate this
  contribution, the average measurement noise squared, $\delta v^2(r)$, from
  the calibration experiment is subtracted from $C_s(s=0,r)$.
\end{thebibliography}

\end{document}